\documentclass[12pt]{article}
 \usepackage{graphicx}
 \usepackage[cp1251]{inputenc} 
\begin{document}
 \noindent {\footnotesize\it Astronomy Letters, 2016, Vol. 42, No. 8, pp. 544--554.}
 \newcommand{\dif}{\textrm{d}}

 \noindent
 \begin{tabular}{llllllllllllllllllllllllllllllllllllllllllllll}
 & & & & & & & & & & & & & & & & & & & & & & & & & & & & & & & & & & & & & \\\hline\hline
 \end{tabular}

  \vskip 0.5cm
  \centerline{\bf\Large The System of Molecular Clouds in the Gould Belt}
  \bigskip
  \centerline{V.V. Bobylev}
  \bigskip
  \centerline{\small\it Pulkovo Astronomical Observatory, St. Petersburg,  Russia}
  \bigskip
  \bigskip
{\bf Abstract}—Based on high-latitude molecular clouds with highly
accurate distance estimates taken from the literature, we have
redetermined the parameters of their spatial orientation. This
system can be approximated by a $350\times235\times140$~pc
ellipsoid inclined by the angle $i=17\pm2^\circ$ to the Galactic
plane with the longitude of the ascending node
$l_\Omega=337\pm1^\circ$. Based on the radial velocities of the
clouds, we have found their group velocity relative to the Sun to
be $(u_0,v_0,w_0)=(10.6,18.2,6.8)\pm(0.9,1.7,1.5)$~km s$^{-1}$.
The trajectory of the center of the molecular cloud system in the
past in a time interval of $\sim$60~Myr has been constructed.
Using data on masers associated with low-mass protostars, we have
calculated the space velocities of the molecular complexes in
Orion, Taurus, Perseus, and Ophiuchus. Their motion in the past is
shown to be not random.


\section*{INTRODUCTION}
A giant stellar-gas complex known as the Gould Belt is located
near the Sun (Frogel and Stothers 1977; Efremov 1989; P\"oppel
1997, 2001; Torra et al. 2000; Olano 2001). A giant cloud of
neutral hydrogen called the Lindblad ring (Lindblad 1967, 2000) is
associated with it; the system of nearby OB associations (de Zeeuw
et al. 1999), open star clusters (Piskunov et al. 2006; Bobylev
2006), and complexes of nearby molecular clouds (Dame et al. 1987;
Perrot and Grenier 2003) are its constituents.

The interest in the Gould Belt stems from the fact that it is the
region of active star formation nearest to the Sun. The star
formation in Orion, Taurus, or Scorpio-Centaurus either has ended
quite recently, as in Scorpio-Centaurus (Preibish and Zinnecker,
1999), or is still continuing, as in the $\rho$ Ophiuchi molecular
cloud (Palla and Staler 2000; Lada 2015). Several projects aimed
at studying the structure and evolution of the Gould Belt and its
components are currently being carried out. These include, for
example, the photometric ``Herschel Gould Belt survey'' (Konyves
et al. 2015) and the ``Gould Belt Very Large Array survey''
devoted to the search for young stars in the Gould Belt (Dzib et
al. 2015). The list is long, because the surveys are being done in
various spectral ranges. Of interest is the astrometric VLBI
survey aimed at determining the trigonometric parallaxes and
proper motions of radio stars in the Gould Belt (Rivera et al.
2015). For a number of stars, the error in the distances
determined by this method is 5\% (Loinard et al. 2008).

The kinematics of the Gould Belt as a whole is quite contradictory
(Palou\v s and Ehlerov\'a 2014; Bobylev 2014). Either expansion or
proper rotation (apart from the Galactic rotation) or both effects
together are found in its motion (Lindblad 2000; Bobylev 2006).
Precisely what is responsible for the formation of the Gould Belt
remains a puzzle. The hypothesis about expansion after a hypernova
explosion proposed by Blaauw (1952) was among the first ones. As
the simulations by Palou\v s (2001) showed, this hypothesis runs
into such difficulties as the absence of a stellar age gradient
along the radius predicted by it or the impossibility to explain
the inclination to the Galactic plane. Nevertheless, a number of
important results were obtained by Olano (1982), Moreno et al.
(1999), P\"oppel and Marronetti (2000), and Perrot and Grenier
(2003) based on the supernova explosion model.

Lepin\'e and Duvert (1994) suggested that the nearby complexes of
molecular clouds located sufficiently far from the Galactic plane
could be formed by the collisions of high-velocity clouds with the
Galactic disk. Since each complex was considered individually, the
Gould Belt within this model could arise only by chance.

Comer\'on and Torra (1992, 1994) considered the more complex model
of an oblique impact of a high-velocity cloud on the Galactic
plane. They showed that a structure similar to the Lindblad ring
but with a considerably larger size is formed as a result. It is
interesting to note the model by Bekki (2009), which is similar to
the Comer\'on--Torra model, but a cloud of dark matter appears
here instead of the high-velocity hydrogen cloud. Bekki's
numerical simulations showed that the Gould Belt could be formed
$\sim$30 Myr ago from an initial gas cloud with a mass of
$\sim10^6~M_\odot$ after its collision with a dark matter clump
with a mass of $3\times10^7~M_\odot$.

The goal of this paper is to refine the spatial orientation
parameters of the system of nearby high-latitude molecular clouds
with known distance estimates. For this purpose, we use the
present-day photometric distances to molecular clouds. To study
their kinematics, we use the radial velocities, trigonometric
parallaxes, and proper motions of radio sources determined by
various authors by VLBI.

\section*{DATA}
In this paper, we use data on 230 molecular clouds from Schlafly
et al. (2014). In the overwhelming majority of cases, these clouds
are within 1~kpc of the Sun. Schlafly et al. (2014) estimated the
distances to them using photometric data on stars from the
Pan-STARRS1 program (Kaiser et al. 2010) in combination with the
distribution map of absorbing matter obtained as a result of the
PLANCK space mission (Planck Collaboration 2011).

The following can be said in more detail about the photometric
method by Schlafly et al. (2014). The distance to each cloud
absorbing the light from the stars behind it was determined in two
steps. In the first step, we determined the reddening and distance
for each individual star in a given direction using Pan-STARRS1
photometry. In the second step, based on stars in the region
selected for our analysis, we constructed the reddening function
as a function of distance with the adoption of a simple dust
screen model. The main idea of the method is to determine the
region of discontinuity between the unreddened foreground stars
and reddened background stars. The position of the discontinuity
region gives an estimate of the distance modulus.

Up to 2000 stars from the region on the celestial sphere with a
radius of 0.2$^\circ$ were used to analyze each cloud. As a
result, according to the compilers of the catalog, the mean random
error in the distances to molecular clouds is less than 5\%,
although the systematic uncertainties resulting from the quality
of the applied stellar models are about 10\%. Thus, this is the
biggest homogeneous modern catalog of accurate, directly measured
distances to molecular clouds. The highly accurate estimates of
the mean distances obtained by these authors for a number of
nearby large molecular complexes, for example, in
Scorpio--Centaurus, Taurus, Perseus, and Orion, are of interest in
studying the Gould Belt.

The radial velocities $V_{LSR}$ are known for quite a few nearby
high-latitude clouds (Magnani et al. 1985). The random errors in
the radial velocities do not exceed 5~km s$^{-1}$, being, on
average, 2~km s$^{-1}$.

Young stars and protostars are closely associated with molecular
clouds. These include, for example, Herbig--Haro objects or
T~Tauri stars. Many of them are observed in the radio band, in
particular, by means of VLBI (Reid et al. 2014). At present, there
are determinations of the radial velocities, trigonometric
parallaxes, and proper motions for a small number of such objects
from the solar neighborhood 0.6 kpc in radius (Loinard 2013; Xu et
al. 2013; Reid et al. 2014). This allows their total space
velocities to be calculated with an error of about 10\%.

VLBI observations of five radio stars in Taurus: Hubble\,4 and
HDE\,283572 (Torres et al. 2007), T\,Tau\,N ($\sim2~M_\odot$;
Loinard et al. 2007), HP\,Tau/G2 (Torres et al. 2009), and V773
Tau (a spectroscopic binary with component masses of
$1.55\pm0.11~M_\odot$ and $1.293\pm0.068~M_\odot$; Torres et al.
2012), were performed at 8.4~GHz in continuum with the VLBA
interferometer in the USA. All these stars are in the neighborhood
of the molecular cloud Lynds 1495; they move parallel to one
another with a small space velocity dispersion.

There are measurements in two clouds with maser sources,
SVC13/NGC1333 (Hirota et al. 2008) and L 1448C (Hirota et al.
2011), in the region of the Perseus molecular complex. VLBI
observations were performed at 22.2~GHz within the Japanese VERA
(VLBI Exploration of Radio Astrometry) program.

In the $\rho$~Ophiuchi molecular cloud, there are measurements for
three stars at the edge close to the Sun: S1 ($\sim6M_\odot$;
Loinard et al. 2008), Dolidze--Arakelyan 21 ($\sim2.2M_\odot$;
Loinard et al. 2008), and YLW~15 (Rivera et al. 2015), as well as
the maser source IRAS 16293$-$2422 at the far edge of the cloud
(Imai et al. 2007). Note that YLW~15 is the only star in our list
without a direct measurement of its trigonometric parallax.

There are two highly accurate results for the Orion molecular
complex. The first is the VLBI measurement of the trigonometric
parallax for the Orion Nebula (Menten et al. 2007) at 8.4 GHz on
the VLBA in the USA. This result was derived here as a mean of the
observations of four different stars (GMR\,A, GMR\,F, GMR\,G, and
GMR\,12). The second highly accurate measurement was made for the
BN/KL object at 43 GHz (SiO maser) within the VERA program (Kim et
al. 2008). As we can see from Fig.~1 in Menten et al. (2007), the
BN/KL object is located at the center of the previous sample of
stars.

Finally, there are VLBI measurements of the trigonometric parallax
and proper motion for the star EC\,95 in Serpens (Dzib et al.
2010). This is a binary system consisting of a Herbig--Haro
protostar with a mass of 4--5~$M_\odot$ and a less massive
T\,Tauri protostar. The systemic radial velocity of this binary
has not yet been determined. Following Honma et al. (2012), we use
$V_{LSR}=9\pm3$~km s$^{-1}$ taken from the velocity map of
molecular clouds (Dame et al. 2001).

 \section*{METHODS}
 \subsection*{Orientation Parameters}
We use the well-known method of determining the symmetry plane of
a stellar system with respect to the main, in our case Galactic,
coordinate system. The basics of this approach were outlined by
Polak (1935); its description can be found in the book by Trumpler
and Weaver (1953). Parenago (1951) and Pavlovskaya (1971)
described the technique for estimating the errors of the
sought-for angles. Recently, this method has been used to
determine the orientation parameters for the system of Cepheids in
the Galaxy (Bobylev 2013) and OB stars belonging to the Gould Belt
(Bobylev and Bajkova 2014a).

In the rectangular coordinate system centered on the Sun, the $x$
axis is directed toward the Galactic center, the $y$ axis is in
the direction of Galactic rotation $(l=90^\circ, b=0^\circ),$ and
the $z$ axis is directed toward the North Galactic Pole; then,
 $x=r\cos l\cos b,$
 $y=r\sin l\cos b,$ and
 $z=r\sin b,$ where $r$ is the
star's heliocentric distance. Let $m,n,k$ be the direction cosines
of the pole of the sought-for great circle from the $x,y,z$ axes.
The sought-for symmetry plane of the stellar system is then
defined as the plane for which the sum of the squares of the
heights, $h=mx+ny+kz,$ is at a minimum:
 \begin{equation}
 \sum h^2=\hbox {min}.
 \label{ff-2}
 \end{equation}
The sum of the squares
 \begin{equation}
 \begin{array}{lll}
 h^2=x^2m^2+y^2n^2+z^2k^2+2yznk+2xzkm+2xymn
 \end{array}
 \label{ff-3}
 \end{equation}
can be designated as $2P=\sum h^2.$ As a result, the problem is
reduced to searching for the minimum of the function $P:$
 \begin{equation}
 \begin{array}{lll}
 2P=am^2+bn^2+ck^2+2fnk+2ekm+2dmn,
 \end{array}
 \label{ff-4}
 \end{equation}
where the second-order moments of the coordinates
 $a=[xx],$
 $b=[yy],$
 $c=[zz],$
 $f=[yz],$
 $e=[xz],$
 $d=[xy],$ written via the Gauss brackets, are the components
of a symmetric tensor:
 \begin{equation}
 \left(\matrix {
  a& d & e\cr
  d& b & f\cr
  e& f & c\cr }\right),
 \label{ff-5}
 \end{equation}
whose eigenvalues $\lambda_{1,2,3}$ are found from the solution of
the secular equation
 \begin{equation}
 \left|\matrix {
a-\lambda&          d&        e\cr
       d & b-\lambda &        f\cr
       e &          f&c-\lambda\cr }\right|=0,
 \label{ff-7}
 \end{equation}
while the directions of the principal axes, $L_{1,2,3}$ and
$B_{1,2,3}$, are found from the relations
 \begin{equation}
 \tan L_{1,2,3}={{ef-(c-\lambda)d}\over{(b-\lambda)(c-\lambda)-f^2}},
 \label{ff-41}
 \end{equation}
 \begin{equation}
 \tan B_{1,2,3}={{(b-\lambda)e-df}\over{f^2-(b-\lambda)(c-\lambda)}}\cos L_{1,2,3}.
 \label{ff-42}
 \end{equation}
The errors in $L_{1,2,3}$ and $B_{1,2,3}$ are estimated according
to the following scheme:
 \begin{equation}
 \varepsilon (L_2)= \varepsilon (L_3)={{\varepsilon (\overline {xy})}\over{a-b}},
 \label{ff-61}
 \end{equation}
 \begin{equation}
 \varepsilon (B_2)= \varepsilon (\varphi)={{\varepsilon (\overline {xz})}\over{a-c}},
 \label{ff-62}
 \end{equation}
 \begin{equation}
 \varepsilon (B_3)= \varepsilon (\psi)={{\varepsilon (\overline {yz})}\over{b-c}},
 \label{ff-63}
 \end{equation}
 \begin{equation}
 \varepsilon^2 (L_1)={\varphi^2\cdot\varepsilon^2(\psi)+\psi^2\cdot\varepsilon^2(\varphi)
   \over{(\varphi^2+\psi^2)^2}},
 \label{ff-64}
 \end{equation}
 \begin{equation}
 \varepsilon^2 (B_1)= {\sin^2 L_1\cdot\varepsilon^2(\psi)+\cos^2L_1\cdot\varepsilon^2(L_1)
   \over{(\sin^2 L_1+\psi^2)^2}},
 \label{ff-65}
 \end{equation}
where
 \begin{equation}
 \varphi=\cot B_1\cdot \cos L_1, \quad \psi=\cot B_1\cdot \sin L_1,
 \label{ff-66}
 \end{equation}
The three quantities
 $\overline {x^2y^2}$, $\overline {x^2z^2}$ and $\overline {y^2z^2}$ should be
calculated in advance. Then,
 \begin{equation}
 \varepsilon^2 (\overline {xy})=(\overline{x^2y^2}-d^2)/n,
 \label{ff-71}
 \end{equation}\begin{equation}
 \varepsilon^2 (\overline {xz})=(\overline {x^2z^2}-e^2)/n,
 \label{ff-72}
 \end{equation}\begin{equation}
 \varepsilon^2 (\overline {yz})=(\overline {y^2z^2}-f^2)/n,
 \label{ff-73}
 \end{equation}
where $n$~--- is the number of stars. Thus, the algorithm for
solving the problem consists in
 (i) setting up the function $2P$~(3),
 (ii) seeking for the roots of the secular equation~(5), and
 (iii) estimating the directions of the principal axes of the
position ellipsoid from Eqs. (6)--(16). As can be seen from
Eqs.~(12), the errors in the directions $L_2$ and $L_3$ coincide,
while the errors in all the remaining directions are calculated
independently of one another.

 \subsection*{Epicyclic Orbits}
Based on the epicyclic approximation (Lindblad 1927), it is easy
to construct the orbits of objects in a coordinate system rotating
around the Galactic center:
 \begin{equation}
 \renewcommand{\arraystretch}{2.2}
 \begin{array}{lll}
 \displaystyle x(t)= x_0+{u_0\over \kappa} \sin(\kappa t)
                     +{v_0\over 2B} (1-\cos(\kappa t)),          \\
 \displaystyle y(t)= y_0+2A \biggl(x_0+{v_0\over 2B}\biggr) t-{\Omega_0 v_0 \over B\kappa} \sin(\kappa t)
                +{2\Omega_0 u_0\over \kappa^2}(1-\cos(\kappa t)),\\
 \displaystyle z(t)= {w_0\over \nu} \sin(\nu t)+ z_0 \cos(\nu t),
 \end{array}
 \label{epiciclic}
 \end{equation}
where $t$ is the time in Myr (we take into account the relation
pc/Myr=0.978 km s$^{-1}$); $A$ and $B$ are the Oort constants;
$\kappa=\sqrt{-4\Omega_0 B}$ is the epicyclic frequency;
$\Omega_0$ is the angular velocity of Galactic rotation of the
local standard of rest; $\Omega_0=A-B$; $\nu=\sqrt{4\pi G\rho_0}$
is the frequency of vertical oscillations, where $G$ is the
gravitational constant and $\rho_0$ is the stellar density in the
solar neighborhood. The object's space velocity components $u,v,w$
are directed along the $x,y,z$ axes, respectively. The parameters
$x_0,y_0,z_0$ and $u_0,v_0,w_0$ in the system of equations (17)
denote the initial positions and velocities of the objects. The
velocities $u,v,w$ are corrected for the peculiar motion of the
Sun relative to the local standard of rest with components
 $(U_\odot,V_\odot,W_\odot)_{LSR}=(11.1,12.2,7.3)$ km s$^{-1}$
(Sch\"onrich et al. 2010). The local density of matter was taken
to be $\rho_0=0.1~M_\odot$~pc$^{-3}$, as estimated by Holmberg and
Flynn (2004), which gives
 $\nu=74$ km s$^{-1}$ kpc$^{-1}$. We used
the Oort constants found from masers with measured trigonometric
parallaxes,
 $A=16.9\pm1.2$ km s$^{-1}$ kpc$^{-1}$ and
 $B=-13.5\pm1.4$ km s$^{-1}$ kpc$^{-1}$ (Stepanishchev and Bobylev 2011),
to which $\kappa= 41$ km s$^{-1}$ kpc$^{-1}$ corresponds. The
Sun's elevation above the Galactic plane is taken to be
$z_0=16\pm2$~pc (Bobylev and Bajkova 2016).

 \begin{figure}
 \begin{center}
 \includegraphics[width=80mm]{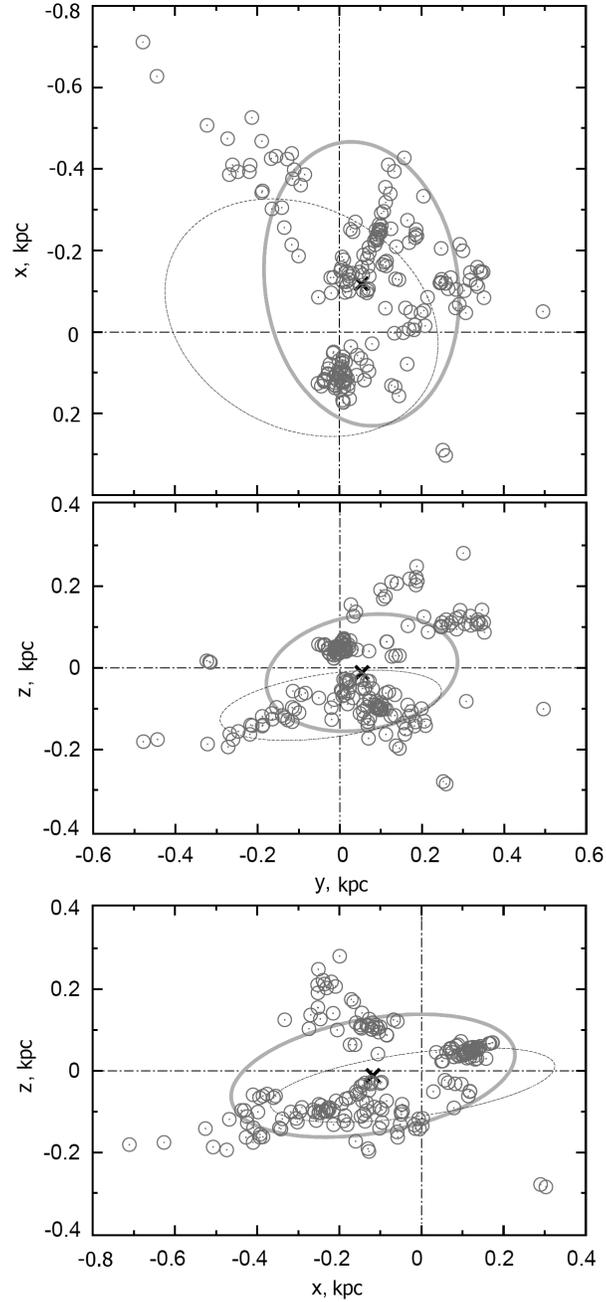}
 \caption{
Positions of 202 molecular clouds in projection onto the Galactic
$xy,$ $yz,$ and $xz$ planes. The Sun is at the coordinate origin.
The thick line indicates the projections of the ellipsoid (18);
the cross marks its center (19). The dashed line indicates the
projections of the ellipsoid found from OB stars (Bobylev and
Bajkova 2014a).
 }
 \label{f-XYZ}
 \end{center}
 \end{figure}
 \begin{figure}
 \begin{center}
 \includegraphics[width=80mm]{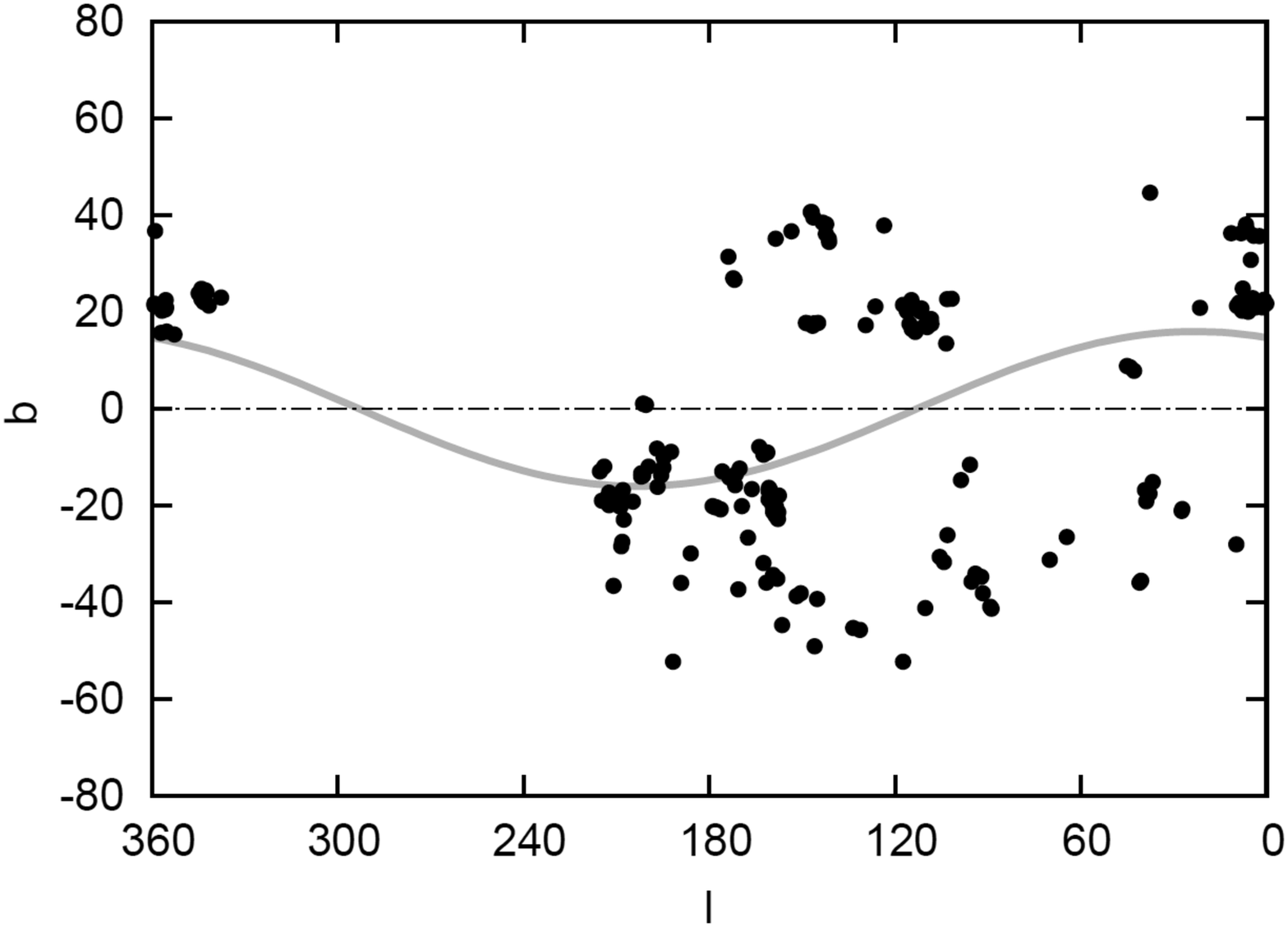}
 \caption{
Distribution of 202 molecular clouds on the celestial sphere. The
sine wave marks the great circle corresponding to the
parameters~(18).
 }
 \label{f-LB}
 \end{center}
 \end{figure}

 \section*{RESULTS AND DISCUSSION}
 \subsection*{Clouds}
The list by Schlafly et al. (2014) contains data for 230 molecular
clouds. We selected 202 clouds located within 1 kpc of the Sun.
Based on this sample, we found the following orientation
parameters:
 \begin{equation}
 \renewcommand{\arraystretch}{1.2}
  \matrix {
  L_1=10\pm15^\circ, & B_1=10\pm15^\circ, \cr
  L_2=103\pm1^\circ, & B_2=15\pm1^\circ, \cr
  L_3=247\pm1^\circ, & B_3=73\pm2^\circ. \cr  }
 \label{rezult-11}
 \end{equation}
The principal semi-axes of the ellipsoid in our method are
determined to within a constant; their ratios are
$\lambda_1:\lambda_2:\lambda_3=1:0.67:0.40.$ If the size of the
first semi-axis is taken to be 350 pc, then the sizes of the
ellipsoid will be $350\times235\times140$~pc, similar to those of
the Lindblad ring. For example, when modelling the evolution of
the neutral hydrogen clouds from which the Gould Belt was formed,
Olano (1982) estimated the size of the system to be
$364\times211$~pc. The coordinates of the ellipsoid center
 \begin{equation}
 \begin{array}{lll}
  x_0=-118\pm15~\hbox{pc}, \\
  y_0= +54\pm11~\hbox{pc}, \\
  z_0= -12\pm 8~\hbox{pc}
 \label{h-11}
 \end{array}
 \end{equation}
show the geometric center of the molecular cloud system to be
located in the second quadrant.

The most important geometric characteristics of the Gould Belt
found by various authors are given in Table 1. These include the
inclination of the system's symmetry plane to the Galactic plane
$i,$ the longitude of the ascending node $l_\Omega$ , and the
rectangular coordinates of the system's geometric center
$x_0,y_0,z_0.$ Stothers and Frogel (1974) studied the distribution
of O--B5 stars. Olano (1982) found the coordinates of the center
of the neutral hydrogen cloud by modelling the consequences of a
supernova explosion. Westin (1985) considered a sample of nearby
stars younger than 60~Myr. Torra et al. (2000) determined the
geometric characteristics of the Gould Belt by analyzing the
distribution of OB stars with their trigonometric parallaxes from
the Hipparcos (1997) catalogue. Bobylev (2004) determined the
geometric characteristics of the Gould Belt by analyzing the
kinematics of OB stars. A sample of OB stars with errors in their
trigonometric parallaxes less than 10\% was also considered by
Bobylev and Bajkova (2014a), who, apart from the parameters listed
in Table~1, also estimated the system's sizes,
$350\times272\times78$~pc (this ellipsoid is shown in Fig. 1).
Perrot and Grenier (2003) studied the distribution and evolution
of molecular clouds; they estimated the sizes of the system to be
$373\times233$~pc. Piskunov et al. (2006) analyzed a sample of
open star clusters, with this sample having been complete within
800~pc of the Sun.

\begin{table}[t]\caption[]{\small
 Geometric characteristics of the Gould Belt found by various authors }
 \begin{center}   \label{t00} \small
 \begin{tabular}{|l|l|l|l|}\hline
      Reference             &          $i$   &      $l_\Omega$ & $(x_0,y_0,z_0)\pm(e_x,e_y,e_z),$ pc\\\hline

 Stothers and Frogel~(1974) & $18\pm1^\circ$ & $295\pm2^\circ$ &          --- \\
 Olano~(1982)               &            --- &             --- & $(-109,126)$ \\
 Westin~(1985)              & $19^\circ$     & $270^\circ$     &          --- \\
 Torra et al.~(2000)        &  $16-22^\circ$ & $275-295^\circ$ &          --- \\
 Perrot and Grenier~(2003)  & $17.2\pm0.5^\circ$ & $296\pm2^\circ$ & $(-104,0)\pm(4,4)$ \\
 Bobylev~(2004)             & $17^\circ$     & $270^\circ$     & $(-141,51)$ \\
 Piskunov et al.~(2006)     &            --- &             --- & $(-78,-53,-46)$ \\
 Bobylev and Bajkova~(2014a) & $13\pm1^\circ$ & $306\pm4^\circ$ & $(-35,-92,-22)\pm(13,14,5)$ \\
 This paper                  & $17\pm2^\circ$ & $337\pm1^\circ$ & $(-118,54,-12)\pm(15,11,8)$ \\
 \hline
 \end{tabular} \end{center}
\end{table}
 \begin{figure}[t]
 \begin{center}
 \includegraphics[width=90mm]{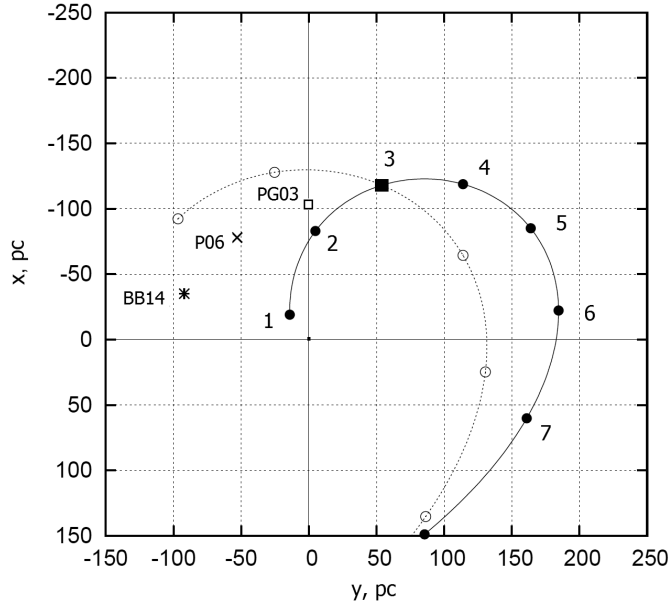}
 \caption{
Trajectory of the center of the molecular cloud system relative to
the local standard of rest (solid line). The filled circles are
placed at intervals of 10 Myr. The filled square (point no. 3)
marks the present-day center of this system. The second trajectory
(the dotted line with open circles) was constructed with a
different peculiar velocity of the Sun (for details, see the
text). P06, BB14, and PG03 designate Piskunov et al. (2006),
Bobylev and Bajkova (2014a), and Perrot and Grenier (2003),
respectively.
 }
 \label{f-GB}
 \end{center}
 \end{figure}

The distributions of molecular clouds in projection onto the
Galactic $xy,yz,xz$ planes are shown in Fig.~1. A significant
difference along the $y$ axis in the position of the ellipse
center found by Bobylev and Bajkova (2014a) from OB stars can be
seen in Fig.~1: $x_0=-35\pm13$~pc, $y_0=-92\pm14$~pc, and
$z_0=-22\pm5$~pc. The situation with the geometric center of the
Gould Belt calculated from the coordinates of open star clusters
by Piskunov et al. (2006) is similar: $x_0=-78$~pc, $y_0=-53$~pc,
and $z_0=-46$~pc. As can be seen from Table~1, the coordinates
(19) are fairly close to those found from gas clouds by Perrot and
Grenier (2003).

Based on the parameters (18), we can calculate two more important
characteristics: the inclination of the symmetry plane of the
molecular cloud system to the Galactic plane $i=17 \pm2^\circ$ and
the longitude of the ascending node
 $l_\Omega=L_3+90^\circ=337\pm1^\circ$.

Note that when compiling their list, Schlafly et al. (2014)
specially relied on high-latitude clouds. This was done, because
it was extremely difficult to separate the Gould Belt clouds at
low latitudes from the Galactic background clouds. In addition,
there are virtually no data in the interval of longitudes
$230^\circ<l<330^\circ$ in the list by Schlafly et al. (2014).
Nevertheless, orientation parameters close to those of the Gould
Belt are reliably obtained from 202 clouds. The distribution of
these clouds on the celestial sphere is shown in Fig.~2, where the
sine wave was constructed with $i=17^\circ$ and
$l_\Omega=337^\circ$ corresponding to the solution~(18). Since the
clouds are high-latitude ones, far from the symmetry plane of the
system (the Gould Belt), the size of the third axis of the
ellipsoid~(18) is approximately twice that of the Gould Belt found
by various authors from stars, OB associations, and open clusters.

The discrepancy between the coordinates of the Gould Belt center
found by various authors from open clusters and from the system of
molecular clouds arouses interest. Is this scatter of results
random? Is a relationship to the age and the individual pattern of
evolution of the samples (star clusters and gas clouds) possible
here?

\begin{table}[t]\caption[]{\small
 Velocities of molecular complexes calculated from radio stars and maser sources }
 \begin{center}      \label{t1} \small
 \begin{tabular}{|l|c|c|c|}\hline
 Complex  & $n$ & $(U,V,W)\pm(e_U,e_V,e_W),$ km s$^{-1}$ & $(x,y,z)\pm(e_x,e_y,e_z),$ pc \\\hline
 TAURUS           & 5 & $(-16.7, -12.4, -9.2)\pm(1.0, 0.2, 0.3)$ & $(-132,~~19,~-40)\pm(~1, 0, 0)$ \\
 ORION            & 5 & $(-16.6, -19.6,~~1.8)\pm(3.8, 2.5, 1.7)$ & $(-343,-190,-137)\pm(4,2,2)$ \\
 OPHIUCHUS        & 3 & $(~-7.0, -14.7, -7.3)\pm(2.1, 0.6, 1.0)$ & $(~114,~-14,~~34)\pm(~6, 1, 2)$ \\
 PERSEUS          & 2 & $(-19.9, -20.7, -6.5)\pm(2.3, 0.7, 0.3)$ & $(-203,~~81,~-83)\pm(15, 6, 6)$ \\
 IRAS 16293--2422 & 1 & $(~-7.9, -31.5, -6.0)\pm(4.4, 6.0, 2.4)$ & $(~171,~-18,~~49)\pm(34, 4, 9)$ \\
 EC 95            & 1 & $(~-3.8, ~-9.1, -5.3)\pm(2.7, 1.7, 0.3)$ & $(~352,~216,~~39)\pm(~3, 2, 0)$ \\\hline
\end{tabular}
\end{center}
\end{table}
 \begin{figure}
 \begin{center}
 \includegraphics[width=90mm]{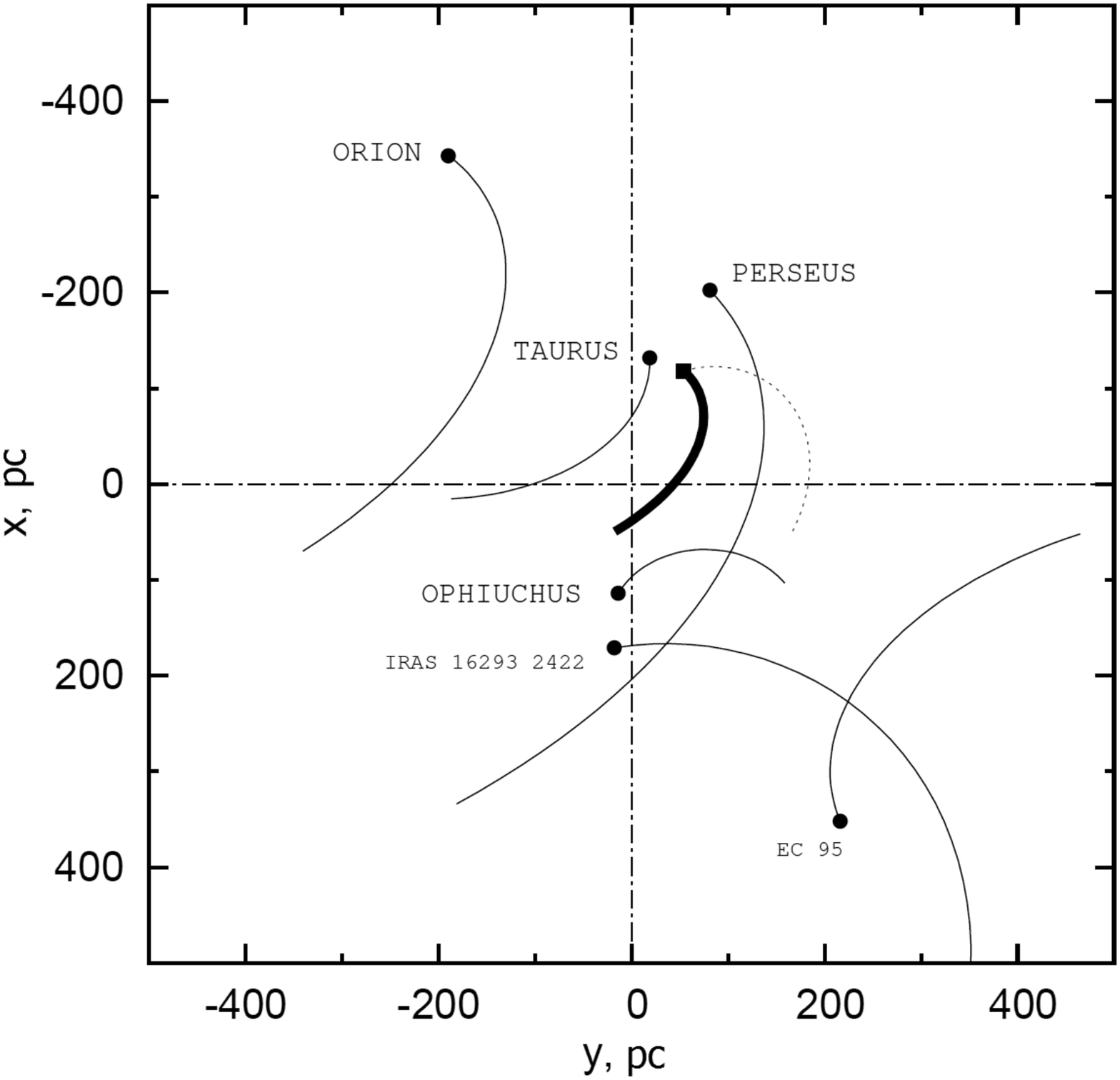}
 \caption{
Trajectories of four molecular complexes and two stars relative to
the local standard of rest in the past in an interval of 40~Myr.
The trajectory of the center of this system is highlighted by the
thick line.
 }
 \label{f-masers14}
 \end{center}
 \end{figure}

First, it is necessary to understand how much we can be in error
when calculating the coordinates of the center. It can be assumed
that the coordinates of the system's center were shifted into the
second quadrant due to the absence of data in the interval of
longitudes $230^\circ<l<330^\circ$. However, it should be noted
that there are no considerable number of molecular clouds
belonging to the Gould Belt in the interval of longitudes
$270^\circ<l<330^\circ$. This can be clearly seen from Fig.~7 in
Dame et al. (1987), where the distribution of molecular complexes
within 1 kpc of the Sun with an indication of their masses and
heights above the Galactic plane are shown. There is a complex of
clouds in Vela only in a direction $l\approx280^\circ$ at a
distance of about 500~pc, but it is located in the Galactic plane
($|b|<25^\circ$). To compensate for the absence of data on
molecular clouds, Perrot and Grenier (2003) took the data on
neutral hydrogen clouds in the region $270^\circ<l<330^\circ$, as
indicated in Figs.~1 and 2 of these authors. Therefore, it is
important that the parameters~(19) are closest to the results of
Perrot and Grenier (2003) obtained by analyzing a sample of gas
clouds distributed quite uniformly in longitude.

Then, we decided to trace the motion of the Gould Belt center in a
time interval comparable to its age (40--60 Myr). For this
purpose, it is necessary to calculate the velocity of the
molecular cloud system relative to the local standard of rest. We
used 105 clouds with the estimates of their radial velocities
$V_{LSR}$ from Magnani et al. (1985). We passed from the
velocities $V_{LSR}$ to $V_r$ via the parameters of the standard
solar motion $(U_\odot,V_\odot,W_\odot)_{LSR}=(10.3,15.3,7.7)$ km
s$^{-1}$. Then,
 \begin{equation}
 \begin{array}{lll}
 V_r=-u_0\cos b\cos l-v_0\cos b\sin l-w_0\sin b+rA\cos^2 b\sin 2l,
 \label{vr-55}
 \end{array}
 \end{equation}
where $A=0.5 R_0\Omega'_0$ is the Oort constant and $\Omega'_0$ is
the first derivative of the angular velocity of Galactic rotation.
Solving the system of conditional equations~(20) by the
least-squares method, we found the following group velocity
components:
 \begin{equation}
 \begin{array}{lll}
  u_0= 10.6\pm0.9~\hbox{km s$^{-1}$}, \\
  v_0= 18.2\pm1.7~\hbox{km s$^{-1}$}, \\
  w_0= ~6.8\pm1.5~\hbox{km s$^{-1}$},
 \label{uvw0}
 \end{array}
 \end{equation}
The Oort constant was $A=13.2\pm6.4$ km s$^{-1}$ kpc$^{-1}$, and
the error per unit weight was $\sigma_0=4.9$ km s$^{-1}$. The
velocities~(21) are in good agreement with the velocity components
of the Gould Belt
$(u_0,v_0,w_0)=(9.26,15.93,6.96)\pm(0.72,0.72,0.60)$ km s$^{-1}$
calculated by Bobylev (2006) using 49 open star clusters. Only the
motion along the y coordinate differs by $\Delta v_0\sim2$ km
s$^{-1}$. Note that the Sun's peculiar motion relative to the
local standard of rest $(U_\odot,V_\odot,W_\odot)_{LSR}$ makes a
major contribution to the group velocity~(21). The residual
velocity is quite low: it is virtually zero along the $z$
coordinate, while it is 4--10 km s$^{-1}$ along each of the $x$
and $y$ coordinates, depending on the adopted solar velocity.

The trajectory of the Gould Belt center relative to the local
standard of rest calculated on the basis of Eqs.~(17) is presented
in Fig.~3. We took the velocities~(21) for its construction. These
velocities were assigned to the determined center of molecular
clouds (point no.~3 in Fig.~3). The segments on the trajectory are
given at intervals of 10~Myr. Numbers 1--7 increase into the past,
i.e., points 1 and 2 are the future ones for point~3.

The solid curve with filled circles in Fig. 3 was calculated with
the velocities corrected for the Sun’s peculiar motion relative to
the local standard of rest with
 $(U_\odot,V_\odot,W_\odot)_{LSR}=(11.1,12.2,7.3)$ km s$^{-1}$ from
Sch\"onrich et al. (2010), while the dotted curve was calculated
with the velocities corrected with
 $(U_\odot,V_\odot,W_\odot)_{LSR}=(6.0,10.6,6.5)$ km s$^{-1}$
 from Bobylev and Bajkova (2014b).
As can be seen from the figure, the trajectory of the center of
the molecular cloud system depends noticeably on the adopted
initial parameters.

Four centers are indicated in Fig.~3: for the system of open star
clusters (P06), the system of OB stars (BB14), the system of
hydrogen clouds (PG03), and the system of molecular clouds (point
no.~3). We get the impression that they are all fall nicely on
some average trajectory of the Gould Belt. But why does the gas
component lag behind the stellar one in this motion? It is
impossible to completely explain this discrepancy by the random
errors in the coordinates of the center.

\subsection*{Complexes of Clouds}
Table 2 provides the velocities of the molecular cloud complexes
calculated from the VLBI measurements of radio stars and maser
sources. Column 1 gives the name of the complex or maser source,
column 2 gives the number of sources that we used to derive the
mean motion, and column 3 and 4 give the mean velocities and mean
coordinates, respectively. As a result, we have four complexes and
two stars.

Figure 4 presents the trajectories of the molecular complexes
relative to the local standard of rest in the past in an interval
of 40 Myr. The trajectory of the center of this system highlighted
by the thick line was calculated for their group velocity
 $(u_0,v_0,w_0)=(13.6,15.0,6.4)\pm(1.7,2.4,1.7)$ km s$^{-1}$ ascribed to
the center of the molecular cloud system (19) using the peculiar
velocity of the Sun relative to the local standard of rest
$(U_\odot,V_\odot,W_\odot)_{LSR}=(11.1,12.2,7.3)$ km s$^{-1}$ from
Sch\"onrich et al. (2010). The dotted line in the figure indicates
the trajectory that is the main trajectory of the center of the
cloud system in the previous figure (Fig.~3).

As can be seen from the figure, the motions of the complexes in
Orion, Taurus, and Perseus are coordinated: these trajectories are
virtually parallel both to one another and to the trajectory of
the center. We may conclude that the random velocity component is
suppressed. In addition, the center indicated in the figure is
fairly close to the geometric center of the sample even 40~Myr
ago. It can be seen from the figure that when each of the
trajectories is directed into the future, an expansion of the
entire structure of the Gould Belt from a point close to its
center is observed.

Using VLBI measurements of radio stars, Rivera et al. (2015)
performed a similar analysis of the kinematics of the complexes in
Orion, Taurus, and Ophiuchus, though without constructing their
trajectories. These authors found the velocities of these
complexes to exhibit an expansion effect. When more such kinematic
data will appear, it will be interesting to trace the spatial
evolution of the system of molecular clouds in the Gould Belt in
details.

\section*{CONCLUSIONS}
Using data on nearby (located within 1 kpc of the Sun) molecular
clouds with highly accurate distance estimates from Schlafly et
al. (2014), we redetermined the spatial orientation parameters of
this system. In particular, the inclination of the symmetry plane
of this system to the Galactic plane and the longitude of the
ascending node were found to be $i=17\pm2^\circ$ and
$l_\Omega=337\pm1^\circ$, respectively. These values show a very
close association of the clouds with the Gould Belt, despite the
gaps in the data, and that the overwhelming majority of clouds are
high-latitude ones. The sizes of the approximation ellipsoid are
$350\times235\times140$~pc, where the third axis is a factor of
2--2.5 larger than the third axis in the ellipsoid of the Gould
Belt found by various authors from young stars, OB associations,
or open star clusters. We showed a great difference between the
coordinates of the Gould Belt center determined by various authors
from open star clusters, OB associations, and gas clouds.

Based on the radial velocities of a sample of 105 high-latitude
clouds, we found their group velocity relative to the Sun
$(u_0,v_0,w_0)=(10.6,18.2,6.8)\pm(0.9,1.7,1.5)$ km s$^{-1}$. Using
these values and based on the epicyclic approximation, we
constructed the trajectory of the center of the system of
molecular clouds in the Gould Belt in an interval of $\sim$60~Myr.

Using highly accurate data on maser sources associated with
low-mass ($1-2 M_\odot$) protostars, we calculated the mean space
velocities of four molecular complexes in Orion, Taurus, Perseus,
and Ophiuchus. We showed the motion of these complexes in the past
to be not random. Such data are still scarce, but invoking the
GAIA measurements will allows us to make considerable progress in
understanding the evolution of the molecular cloud system and the
Gould Belt as a whole.

\subsection*{ACKNOWLEDGMENTS}
We are grateful to the referees for their helpful remarks that
contributed to an improvement of this paper. This work was
supported by the ``Transitional and Explosive Processes in
Astrophysics'' Program P--41 of the Presidium of Russian Academy
of Sciences.

 \bigskip{REFERENCES}\medskip
 {\small

1. P.A.R. Ade, N. Aghanim, M. Arnaud, M. Ashdown, J. Aumont, C.
Baccigalupi, A. Balbi, A.J. Banday, et al. (Planck Collab.),
Astron. Astrophys. 536, A19 (2011).

2. K. Bekki, Mon. Not. R. Astron. Soc. 398, L36 (2009).

3. A. Blaauw, Bull. Astron. Inst. Netherlands 11, 414 (1952).

4. V.V. Bobylev, Astron. Lett. 30, 159 (2004).

5. V. V. Bobylev, Astron. Lett. 32, 816 (2006).

6. V.V. Bobylev, Astron. Lett. 39, 753 (2013).

7. V.V. Bobylev and A.T. Bajkova, Astron. Lett. 40, 783 (2014).

8. V.V. Bobylev, Astrophysics 57, 583 (2014).

9. V.V. Bobylev, and A.T. Bajkova, Mon. Not. R. Astron. Soc. 441,
142 (2014b).

10. V.V. Bobylev and A.T. Bajkova, Astron. Lett. 42, 1 (2016).

11. F. Comer\'on and J. Torra, Astron. Astrophys. 261, 94 (1992).

12. F. Comer\'on and J. Torra, Astron. Astrophys. 281, 354 (1994).

13. T.M. Dame, H. Ungerechts, R.S. Cohen, E.J. de Geus, I.A.
Grenier, J. May, D.C. Murphy, L.-A. Nyman, and P. Thaddeus,
Astrophys. J. 322, 706 (1987).

14. T.M. Dame, D. Hartmann, and P. Thaddeus, Astrophys. J. 547,
792 (2001).

15. S. Dzib, L. Loinard, A.J. Mioduszewski, A.F. Boden, L.F.
Rodr\'iguez, and R.M. Torres,Astrophys. J. 718, 610 (2010).

16. S. Dzib, L. Loinard, L.F. Rodr\'iguez, A.J. Mioduszewski, G.N.
Ortiz-Leon, M.A. Kounkel, G. Pech, J.L. Rivera, et al., Astrophys.
J. 801, 91 (2015).

17. Yu.N. Efremov, Sites of Star Formation in Galaxies (Nauka,
Moscow, 1989) [in Russian].

18. J.A. Frogel and R. Stothers, Astron. J. 82, 890 (1977).

19. The Hipparcos and Tycho Catalogues, ESA SP-1200 (1997).

20. T. Hirota, T. Bushimata, Y.K. Choi, M. Honma, H. Imai, I.
Hiroshi, K. Iwadate, T. Jike, et al., Publ. Astron. Soc. Jpn. 60,
37 (2008).

21. T. Hirota, M. Honma, H. Imai, K. Sunada, Y. Ueno, H.
Kobayashi, and N. Kawaguchi, Publ. Astron. Soc. Jpn. 63, 1 (2011).

22. J. Holmberg and C. Flynn, Mon. Not. R. Astron. Soc. 352, 440
(2004).

23. M. Honma, T. Nagayama, K. Ando, T. Bushimata, Y.K. Choi, T.
Handa, T. Hirota, H. Imai, et al., Publ. Astron. Soc. Jpn. 64, 136
(2012).

24. H. Imai, K. Nakashima, T. Bushimata, Y.K. Choi, T. Hirota, M.
Honma, K. Horiai, N. Inomata, et al., Publ. Astron. Soc. Jpn. 59,
1107 (2007).

25. M.K. Kim, T. Hirota, M. Honma, H. Kobayashi, T. Bushimata,
Y.K. Choi, H. Imai, K. Iwadate, et al., Publ. Astron. Soc. Jpn.
60, 991 (2008).

26. V. Konyves, Ph. Andre, A. Men’shchikov, P. Palmeirim, D.
Arzoumanian, N. Schneider, A. Roy, P. Didelon, et al., Astron.
Astrophys. 584, 91 (2015).

27. C.J. Lada, in Young Stars and Planets Near the Sun,
Proceedings of the IAU Symposium No. 314, Ed. by J.H. Kastner, B.
Stelzer, and S.A. Metchev (2015).

28. J.R.D. Lepin\'e and G. Duvert, Astron. Astrophys. 286, 60
(1994).

29. B. Lindblad, Ark. Mat. Astron. Fys. A 20 (17) (1927).

30. P.O. Lindblad, Bull. Astron. Inst. Netherland 19, 34 (1967).

31. P.O. Lindblad, Astron. Astrophys. 363, 154 (2000).

32. L. Loinard, R.M. Torres, A.J. Mioduszewski, L.F. Rodriguez,
R.A. Gonzalez-Lopezlira, R. Lachaume, V. Vazquez, and E. Gonzalez,
Astrophys. J. 671, 546 (2007).

33. L. Loinard, R.M. Torres, A.J. Mioduszewski, and L.F.
Rodriguez, Astrophys. J. 675, L29 (2008).

34. L. Loinard, in Proceedings of the IAU Symposium No. 289 on
Advancing the Physics of Cosmic Distances, Beijing, China, Aug.
27–31, 2012, Ed. by R. de Grijs and G. Bono (2013).

35. L. Magnani, L. Blitz, and L. Mundy, Astrophys. J. 295, 402
(1985).

36. K.M. Menten, M.J. Reid, J. Forbrich, and A. Brunthaler,
Astron. Astrophys. 474, 515 (2007).

37. E. Moreno, E.J. Alfaro, and J. Franco, Astrophys. J. 522, 276
(1999).

38. C.A. Olano, Astron. Astrophys. 112, 195 (1982).

39. C.A. Olano, Astron. Astrophys. 121, 295 (2001).

40. F. Palla and S. W. Staler, Astrophys. J. 540, 255 (2000).

41. J. Palou\v s, Astrophys. Space Sci. 276, 359 (2001).

42. J. Palou\v s, and S. Ehlerov\'a, arXiv:1406.6248 (2014).

43. P.P. Parenago, Tr. GAISh 20, 26 (1951).

44. E.D. Pavlovskaya, in Practical Works on Stellar Astronomy, Ed.
by P.G. Kulikovskii (Nauka, Moscow, 1971), p. 162 [in Russian].

45. C.A. Perrot and I.A. Grenier, Astron. Astrophys. 404, 519
(2003).

46. A.E. Piskunov, N.V. Kharchenko, S. R\"oser, E. Schilbach, and
R.-D. Scholz, Astron. Astrophys. 445, 545 (2006).

47. I.F. Polak, Introduction to Stellar Astronomy (ONTI, Moscow,
Leningrad, 1935) [in Russian].

48. W.G.L. P\"oppel, Fundam. Cosm. Phys. 18, 1 (1997).

49. W.G.L. P\"oppel, ASP Conf. Ser. 243, 667 (2001).

50. W.G.L. P\"oppel and P. Marronetti, Astron. Astrophys. 358, 299
(2000).

51. T. Preibish and H. Zinnecker, Astron. J. 117, 2381 (1999).

52. M.J. Reid, K.M. Menten, A. Brunthaler, X.W. Zheng, T.M. Dame,
Y. Xu, Y. Wu, B. Zhang, et al., Astrophys. J. 783, 130 (2014).

53. J.L. Rivera, L. Loinard, S.A. Dzib, G.N. Ortiz-Leon, L.F.
Rodriguez, and R.M. Torres, Astrophys. J. 807, 119 (2015).

54. E.F. Schlafly, G. Green, D.P. Finkbeiner, H.- W. Rix, E.F.
Bell, W.S. Burgett, K.C. Chambers, P.W. Draper, et al., Astrophys.
J. 786, 29 (2014).

55. R. Sch\"onrich, J. Binney, and W. Dehnen, Mon. Not. R. Astron.
Soc. 403, 1829 (2010).

56. A.S. Stepanishchev and V.V. Bobylev, Astron. Lett. 37, 254
(2011).

57. R. Stothers and J.A. Frogel, Astron. J. 79, 456 (1974).

58. J. Torra, D. Ferna\'ndez, and F. Figueras, Astron. Astrophys.
359, 82 (2000).

59. R.M. Torres, L. Loinard, A.J. Mioduszewski, and L.F.
Rodriguez, Astrophys. J. 671, 1813 (2007).

60. R.M. Torres, L. Loinard, A.J. Mioduszewski, and L.F.
Rodriguez, Astrophys. J. 698, 242 (2009).

61. R.M. Torres, L. Loinard, A.J. Mioduszewski, A.F. Boden, R.
Franco-Hernandez, W.H.T. Vlemmings, and L.F. Rodriguez, Astrophys.
J. 747, 18 (2012).

62. R.J. Trumpler and H.F. Weaver, Statistical Astronomy (Univ.
California Press, Berkely, 1953).

63. T.N.G. Westin, Astron. Astrophys. Suppl. Ser. 60, 99 (1985).

64. Y. Xu, J.J. Li, M.J. Reid, K.M. Menten, X.W. Zheng, A.
Brunthaler, L. Moscadelli, T.M. Dame, and B. Zhang, Astrophys. J.
769, 15 (2013).

65. P.T. de Zeeuw, R. Hoogerwerf, J.H.J. de Bruijne, A.G.A. Brown,
and A. Blaauw, Astron. J. 117, 354 (1999).

 }

\end{document}